\documentclass[sigconf,nonacm]{acmart}

\AtBeginDocument{%
  }
\usepackage{booktabs}
\usepackage{xcolor}

\providecolor{USTARBlue}{HTML}{2878B5}
\providecolor{USTARRed}{HTML}{C43C39}
\usepackage{booktabs}
\usepackage{xcolor}

\providecolor{USTARBlue}{HTML}{2878B5}
\providecolor{USTAROrange}{HTML}{B66A16}
\providecolor{USTARGreen}{HTML}{287A4B}
\providecolor{USTARPurple}{HTML}{7851A9}

\usepackage{booktabs}
\usepackage{xcolor}
\usepackage{colortbl}
\usepackage{tikz}
\usepackage{pgfplots}
\usepgfplotslibrary{groupplots}

\pgfplotsset{compat=1.18}

\definecolor{SIDBlue}{HTML}{1663A6}
\definecolor{SIDYellow}{HTML}{D9A321}

\pgfplotsset{
    sidaxis/.style={
        width=0.44\textwidth,
        height=5.4cm,
        font=\small,
        xmin=0.8,
        xmax=3.2,
        xtick={1,2,3},
        xticklabels={
            {Level 1},
            {Level 2},
            {Level 3}
        },
        tick label style={font=\footnotesize},
        label style={font=\small},
        title style={font=\small},
        axis line style={black!65},
        tick style={black!65},
        tick align=outside,
        grid=major,
        grid style={
            draw=gray!20,
            line width=0.3pt
        },
        legend style={
            draw=none,
            fill=white,
            fill opacity=0.9,
            text opacity=1,
            font=\footnotesize,
            cells={anchor=west},
            row sep=2pt
        }
    },
    baselinecurve/.style={
        color=SIDYellow,
        line width=1.5pt,
        mark=square*,
        mark size=2.8pt,
        mark options={
            solid,
            fill=SIDYellow,
            draw=SIDYellow
        }
    },
    ustarcurve/.style={
        color=SIDBlue,
        line width=1.5pt,
        mark=*,
        mark size=2.8pt,
        mark options={
            solid,
            fill=SIDBlue,
            draw=SIDBlue
        }
    }
}
\definecolor{USTARBlue}{HTML}{1663A6}
\definecolor{USTARLightBlue}{HTML}{EEF5FC}

\usepackage{algorithm}
\usepackage{algpseudocode}
\usepackage{xcolor}
\usepackage{colortbl}  
\usepackage{tabularx}
\usepackage{array}
\usepackage[table]{xcolor}
\definecolor{USTARBlue}{HTML}{1663A6}
\definecolor{USTARYellow}{HTML}{B8860B}
\definecolor{USTARGreen}{HTML}{2E7D32}
\definecolor{USTARRed}{HTML}{C62828}

\definecolor{USTARLightBlue}{HTML}{EEF5FC}
\definecolor{USTARLightYellow}{HTML}{FFF8E1}
\definecolor{USTARLightGreen}{HTML}{EAF5EB}
\definecolor{USTARLightRed}{HTML}{FDECEC}
\pgfplotsset{
    single bar legend/.style={
        legend image code/.code={
            \draw[##1] (0cm,-0.08cm) rectangle (0.22cm,0.08cm);
        }
    }
}
\begin{document}
\pagestyle{plain}
\title{SPAR: Stepwise Personalized Autoregressive Recommendation
with Prefix-Balanced Semantic IDs}

\author{Chunrui Zhang}
\email{zhangcr127@gmail.com}
\affiliation{
  \institution{University of Electronic Science and Technology of China}
  \city{Chengdu}
  \country{China}
}

\author{Gaoyun Lin}
\email{gaoyunlin@mail.ustc.edu.cn}
\affiliation{
  \institution{University of Science and Technology of China}
  \city{Hefei}
  \country{China}
}
\author{Chang Zhou}
\email{changzhou.zc@alibaba-inc.com}
\affiliation{
  \institution{HUJING Digital Media \& Entertainment Group}
  \city{Beijing}
  \country{China}
}
\author{Zheng Hu}
\authornote{Corresponding author}
\email{huzheng@swjtu.edu.cn}
\affiliation{
  \institution{Southwest Jiaotong University}
  \city{Chengdu}
  \country{China}
}
\author{Qiran Song}
\email{qrsong0907@163.com}
\affiliation{
  \institution{University of Electronic Science and Technology of China}
  \city{Chengdu}
  \country{China}
}
\author{Shimin Cai}
\email{shimincai@uestc.edu.cn}
\affiliation{
  \institution{University of Electronic Science and Technology of China}
  \city{Chengdu}
  \country{China}
}

\author{Tao Zhou}
\email{zhutou@ustc.edu.cn}
\affiliation{
  \institution{University of Electronic Science and Technology of China}
  \city{Chengdu}
  \country{China}
}
\renewcommand{\shortauthors}{Zhang et al.}

\begin{abstract}
Semantic ID (SID)-based generative recommendation decomposes item selection into a sequence of token-level decisions, making
personalization at each decoding step essential. However, each generated prefix defines a different candidate set, while existing methods insufficiently account for prefix-dependent user evidence and imbalanced historical interaction loads across candidate branches. These limitations increase the risk of irreversible branch-selection errors and reinforce the dominance of popular paths during autoregressive decoding. To address these challenges, we propose SPAR, a generative recommendation framework that enables stepwise personalized branch selection by integrating prefix-conditioned user evidence with interaction-load-aware SID construction. Technically, we design PATH Decoder, a Personalized Adaptive Tree-aware Hybrid Decoder that dynamically retrieves prefix-relevant evidence from compressed long-term memories and integrates it with recent behavior, user profiles, and event context for stepwise personalized decoding. Complementarily, we introduce RQ-PIB, a Prefix-aware Interaction-Balanced SID construction method that regulates historical interaction loads among sibling branches under each parent prefix, mitigating popularity-driven imbalance in the SID hierarchy. We conduct extensive offline and online experiments on Youku,
a major video streaming platform operated by Alibaba.
Offline evaluations on large-scale industrial data demonstrate
an average relative improvement of 2.66\% in HitRate over
the strongest baseline. Further analysis shows that SPAR
improves target-prefix retention by an average of 5.79 percentage points and reduces the traffic-weighted largest
child-branch share by an average of 32.84\% across SID levels. A 7-day online A/B test deploying SPAR to 10\% of Youku's production traffic further achieves relative improvements of 3.95\% in average watch time per user and 3.38\% in play conversion rate. These results demonstrate the effectiveness of SPAR in improving hierarchical decoding reliability and mitigating
popularity-driven branch imbalance, ultimately enhancing
retrieval performance and user engagement in real-world
recommendation scenarios.
\end{abstract}



\maketitle

\begin{figure}[t]
    \centering
    \includegraphics[width=\columnwidth]{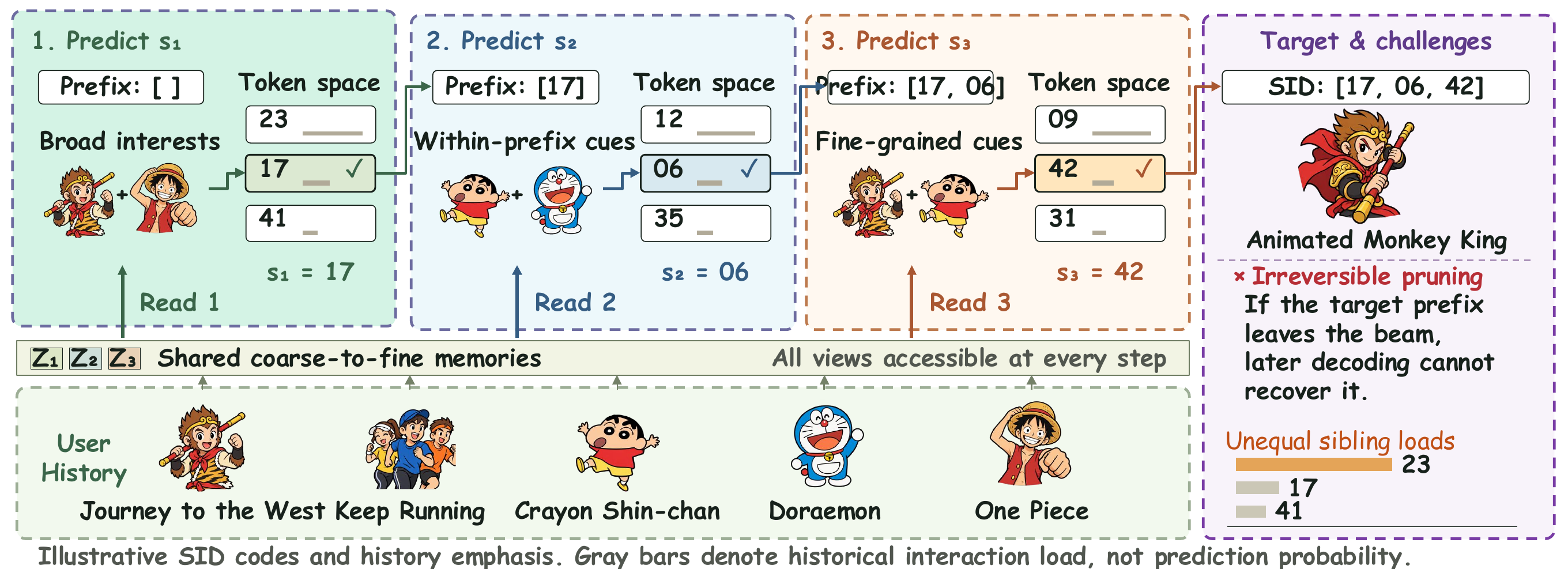}
    \caption{An example of prefix-conditioned interest extraction in hierarchical decoding. As the generated SID prefix changes, each decoding step requires different evidence from the same long-term user history to distinguish the remaining candidate branches. This motivates prefix-conditioned interest extraction throughout the decoding process. The example also illustrates the challenges of irreversible target-prefix pruning and imbalanced historical interaction loads among sibling branches.}
    \label{fig:problem}
\end{figure}
\section{Introduction}

Web platforms such as video streaming services, e-commerce sites,
and social feeds rely on large-scale recommendation to connect
users with relevant content under strict latency constraints.
Generative retrieval reformulates this problem as autoregressive
item-identifier generation~\cite{ripor,ed2}. In Semantic ID
(SID)-based methods, each item is represented by a sequence of
discrete semantic tokens, enabling retrieval through successive
token predictions conditioned on user behavior~\cite{ed2}.

Existing studies have advanced SID-based generative recommendation
along two directions. The first focuses on item identifier
construction, including semantic tokenization~\cite{kdd},
load-regulated code assignment~\cite{unirec}, and hierarchical
category-tree representations~\cite{categorytree}.
The second improves user-conditioned generation through
generative architectures~\cite{onerec,onerecv2} and contextual
information access~\cite{cllm4rec}. Despite these advances,
existing methods primarily optimize item representations or
generation architectures, without explicitly coordinating
prefix-dependent user evidence with the organization of
candidate branches. This leaves the hierarchical nature
of SID decoding insufficiently addressed. More importantly,
the two research directions are often considered separately,
although they jointly determine the outcome of autoregressive
retrieval. SID construction defines the candidate branches
available at each prefix, while the generation model determines
how user preferences are translated into branch probabilities.
Improving either component without considering their interaction
may leave a mismatch between the structure of the candidate
space and the personalized evidence used to navigate it.

The key challenge lies in the changing requirements of
prefix-conditioned branch decisions. Near the root of the
SID tree, the decoder requires coarse-grained interest
evidence, while deeper decisions demand finer information
to distinguish sibling branches.
As illustrated in Figure~\ref{fig:problem}, generating
the target SID $[17,06,42]$ requires progressively shifting
attention from broad user interests to prefix-specific
preferences and fine-grained behavioral cues as the
candidate space narrows.
For users with diverse and evolving preferences, a fixed
user representation or static long-term summary~\cite{rella,retr}
may fail to capture the evidence most relevant to each decision.
For instance, a user may exhibit broad interests in
multiple content categories, while individual branches
within each category require more specific behavioral
signals for discrimination. Evidence useful for selecting
a coarse semantic category may therefore become less
informative once the decoder enters a particular subtree.
Conversely, fine-grained preferences that are irrelevant
to the initial decision may become decisive at deeper
levels. This creates a dynamic evidence-selection problem
that cannot be fully addressed by simply increasing
the capacity of a shared user representation. Instead,
personalization should explicitly account for the
current generated prefix and its remaining candidate
branches.

Meanwhile, uneven historical interaction loads introduce
another difficulty. As also illustrated in Figure~\ref{fig:problem},
popular SID paths attract disproportionately more interactions,
potentially dominating branch selection and irreversibly
eliminating less frequent but relevant targets.
Since branch competition occurs among siblings under the
same prefix, global codeword balancing may overlook
such local disparities. In particular, the same codeword
can appear under different parent prefixes with distinct
conditional branch distributions, allowing balanced
aggregate loads to conceal highly skewed sibling loads.
Moreover, enforcing uniform branch assignments may disrupt
semantic proximity, introducing a trade-off between load
balance and representation quality. Therefore, effective
SID-based retrieval requires both prefix-dependent user
evidence and locally balanced candidate branches while
preserving semantic structure.

To address these challenges, we propose SPAR, a Stepwise
Personalized Autoregressive Recommendation framework with
Prefix-Balanced Semantic IDs. SPAR integrates two complementary
components. First, PATH Decoder constructs coarse-to-fine
long-term history views and compresses them into compact
memory banks. A decoder-only backbone models recent SID
interactions, while prefix-conditioned cross-attention
dynamically retrieves relevant long-term evidence at each
generation step. User and Event Tokens further incorporate
profile information and request-time context into token
prediction. Second, RQ-PIB constructs semantic identifiers
through residual quantization and rebalances historical
interaction loads among child branches sharing the same
parent prefix. By selectively reassigning items with minimal
additional semantic distortion, it mitigates local branch
dominance while preserving semantic structure. Together,
PATH Decoder and RQ-PIB improve both the evidence used
for branch selection and the organization of competing
branches. SPAR further incorporates preference-aligned
training to refine generated recommendations.

Our main contributions are summarized as follows:

\begin{itemize}
    \item \textbf{Stepwise Personalized Decoding.}
    We propose PATH Decoder, which dynamically retrieves
    prefix-relevant evidence from compact coarse-to-fine
    long-term memories and integrates recent behavior,
    user profiles, and event context for personalized
    token prediction.

    \item \textbf{Prefix-Balanced SID Construction.}
    We introduce RQ-PIB to regulate historical interaction
    loads among sibling branches under the same prefix
    while limiting additional semantic reconstruction
    error, addressing local imbalances overlooked by
    global codeword balancing.

    \item \textbf{Large-Scale Empirical Evaluation.}
    Experiments on industrial Youku data demonstrate an
    average relative improvement of 2.66\% across
    HR@10/20/50/100 over the strongest baseline.
    Online A/B tests further improve average watch time
    per user by 3.95\% and play conversion rate by 3.38\%.
\end{itemize}
\section{Related Work}

\subsection{\texorpdfstring{Generative Recommendation}{Generative Recommendation}}

Traditional large-scale recommendation systems typically adopt a
multi-stage retrieval architecture, where two-tower models such as
YouTubeDNN~\cite{youtubednn} efficiently retrieve candidates through
user--item embedding matching. Despite their efficiency, these
methods rely on predefined item representations and similarity-based
retrieval, limiting the expressiveness of user--item interactions.
Generative recommendation (GR) offers an alternative paradigm by
directly generating recommendation targets. Early efforts such as
P5~\cite{geng2022p5} formulate recommendation as personalized text
generation, while HSTU~\cite{zhai2024hstu} explores scalable
sequential modeling of user behavior. More recently, Semantic ID
(SID)-based generative retrieval has emerged as a promising approach,
representing items as sequences of discrete tokens and retrieving
candidates through autoregressive generation rather than conventional
embedding matching.

Existing SID-based methods primarily focus on item tokenization
and generative modeling. TIGER~\cite{tiger} introduces residual
quantization for semantic item indexing and autoregressive retrieval;
CoST~\cite{zhu2024cost} enhances semantic consistency through
contrastive quantization; LETTER~\cite{wang2024letter} incorporates
semantic, collaborative, and diversity objectives into item
tokenization; and RPG~\cite{hou2025generating} improves generation
efficiency through parallel SID decoding. Industrial frameworks,
including UniRec~\cite{unirec}, OneRec~\cite{onerec}, and
OneRec-V2~\cite{onerecv2}, further advance generative recommendation
through scalable architectures and unified modeling. However, these
approaches largely emphasize the construction of item identifiers
or the optimization of generation frameworks, while paying limited
attention to the hierarchical decision structure induced by SID
prefixes. In particular, uneven branch loads and evolving user
preferences across decoding steps can undermine retrieval
effectiveness. In contrast, our work explicitly addresses these
challenges through prefix-conditioned branch balancing and stepwise
personalized decoding, jointly improving SID organization and
user-aware decision-making at each level of the generation tree.
\subsection{\texorpdfstring{User Interest Modeling and Personalization}{User Interest Modeling and Personalization}}

User interest modeling is fundamental to personalized recommendation.
Early sequential methods, including GRU4Rec~\cite{hidasi2016gru4rec},
SASRec~\cite{kang2018sasrec}, and BERT4Rec~\cite{sun2019bert4rec},
capture evolving user preferences through recurrent networks and
attention-based architectures, while S3-Rec~\cite{zhou2020s3rec}
further enhances sequential representations through self-supervised
learning. Subsequent methods such as UniSRec~\cite{hou2022towards}
and RecFormer~\cite{li2023text} incorporate textual semantics to
improve generalization, whereas LLM-based approaches, including
TALLRec~\cite{bao2023tallrec}, LC-Rec~\cite{zheng2024adapting},
and CALRec~\cite{li2024calrec}, explore language-model adaptation
for personalized recommendation. Beyond sequential modeling,
industrial systems emphasize fine-grained interest extraction:
MIND~\cite{li2019mind} captures multiple user interests,
DIN~\cite{zhou2018din} and DIEN~\cite{zhou2019dien} introduce
target-aware behavior modeling, and MIMN~\cite{pi2019mimn}
and SIM~\cite{pi2020sim} extend interest modeling to long behavior
sequences through memory and retrieval mechanisms.

Despite these advances, existing methods primarily model user
preferences through sequence-level representations or target-aware
aggregation, where the candidate item is explicitly available.
In contrast, SID-based generative retrieval makes a sequence of
hierarchical branch decisions, with only a partial target prefix
available at each decoding step. Consequently, the relevant user
evidence may vary across SID levels and candidate branches, making
static user representations or conventional target-aware attention
insufficient for fine-grained personalization. To address this
distinction, SPAR organizes long-term user behaviors into compact
memories at multiple SID-tree depths and models recent interactions
through the autoregressive backbone. More importantly, it dynamically
retrieves relevant historical evidence conditioned on the current
SID prefix, enabling stepwise personalization at the branch-decision
level rather than relying solely on sequence-level user representations.
\section{Method}
\begin{figure*}[t]
    \centering
    \includegraphics[width=\textwidth]{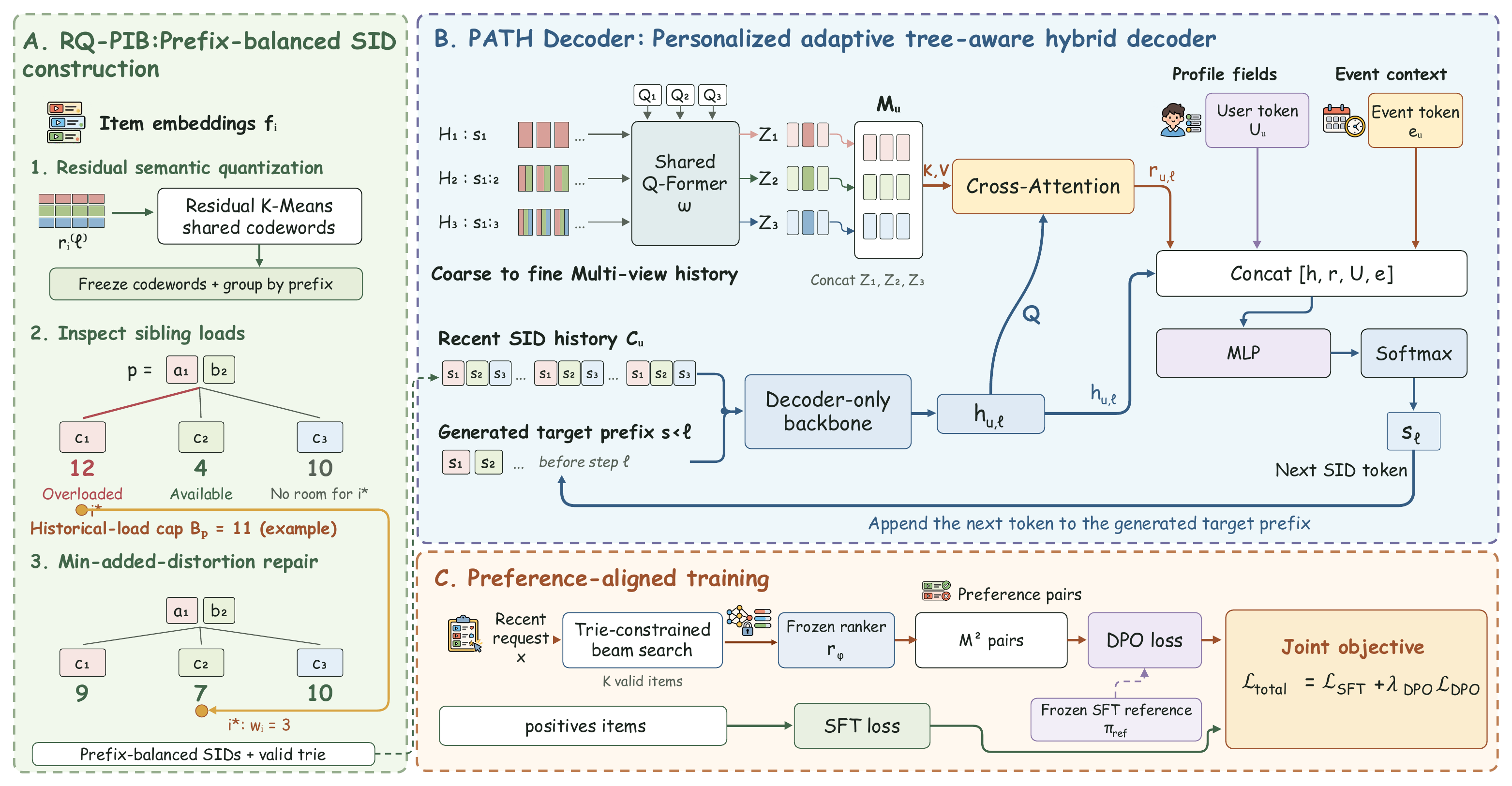}
    \caption{Overall framework of SPAR. (A) RQ-PIB regulates historical interaction load among sibling branches under the same parent prefix while limiting additional semantic distortion. (B) PATH Decoder retrieves relevant information from coarse-to-fine long-term user memories according to the current decoding state and generated prefix, and integrates it with recent behavior, user profiles, and event context for next-token prediction. (C) Preference-aligned training constructs DPO pairs from beam-search candidates using a reward model and jointly optimizes the  supervised fine-tuning (SFT) and DPO objectives.}
    \label{fig:framework}
\end{figure*}
\subsection{Overview}
\label{sec:overview}
SPAR consists of three stages, as illustrated in
Figure~\ref{fig:framework}: RQ-PIB constructs prefix-balanced SIDs
offline; PATH Decoder retrieves prefix-relevant long-term evidence
and fuses it with recent behavior, profile, and event context for
next-token prediction; and preference alignment adapts the
generator to recent request-time preferences.

\subsection{PATH Decoder: Personalized Adaptive
Tree-aware Hybrid Decoder}
To support stepwise personalized decoding, we design the PATH Decoder, which consists of two key components: (1) a sequence-evidence module that integrates long-term tree-view memories with recent interactions through prefix-conditioned memory retrieval, and (2) hybrid decision tokens that incorporate user-profile and event context into the decoding process.

\textbf{Long-term Coarse-to-fine history views.}
Let the long-term history be \(\mathcal H_u^{L}=[v_{u,1},\ldots,v_{u,n}]\), ordered chronologically. We construct three history views from cumulative SID prefixes, where $k \in \{1,\ldots,N_{\mathrm{SID}}\}$ denotes the SID layer,
and $N_{\mathrm{SID}}$ is the total number of SID layers:
\begin{equation}
\label{eq:ustar_history_views}
\begin{aligned}
\mathbf x_{u,j}^{(k)}
&=\sum_{r=1}^{k}\mathbf E_r[s_r(v_{u,j})]
 +\mathbf e_{\mathrm{pos}}(j)
 +\mathbf e_{\mathrm{level}}(k),\\
\mathbf H_{u,k}
&=\left[
\mathbf x_{u,1}^{(k)};\ldots;\mathbf x_{u,n}^{(k)}
\right].
\end{aligned}
\end{equation}
Here, \(\mathbf E_r\) is an SID embedding table, while the position and level embeddings identify interaction order and SID depth. The sequences \(\mathbf H_{u,1},\mathbf H_{u,2},\mathbf H_{u,3}\) retain one-, two-, and three-token prefixes, respectively, capturing user interests at progressively finer SID resolutions. These views align historical interactions with different depths of
the SID tree, so the decoder can access coarse interests near the
root and fine-grained preferences near deeper prefix nodes.

\textbf{Long-term memory compression.}
Each history view is compressed with a Q-Former-style encoder~ \cite{li2023blip}. We use a separate bank of learnable queries \(\mathbf Q_k\) for each view and share the encoder parameters \(\omega\):
\begin{equation}
\label{eq:ustar_sequence_compression}
\mathbf Z_{u,k}
=\operatorname{QFormer}_{\omega}
\left(\mathbf Q_k,\mathbf H_{u,k}\right).
\end{equation}
The shared encoder extracts behavioral patterns across resolutions, while the view-specific queries produce compact memories that retain their respective hierarchical information.

\textbf{Recent SID sequence backbone.}
The decoder-only backbone receives the recent interaction history as a sequence of SID tokens. Let \(m\) denote the number of recent interactions included in the decoder context, and let \( v_{u,j}\) be the \(j\)-th recent item, its input block and the resulting backbone context are
\begin{equation}
\label{eq:ustar_sid_context}
\begin{aligned}
\mathbf B_{u,j}
&=\left[
\mathbf E_1[s_1(v_{u,j})];
\mathbf E_2[s_2(v_{u,j})];
\mathbf E_3[s_3(v_{u,j})]
\right],\\
\mathbf C_u
&=\left[
\mathbf B_{u,1};\ldots;\mathbf B_{u,m}
\right].
\end{aligned}
\end{equation}
All interactions precede the target item. Before predicting SID position \(\ell\), the backbone computes
\begin{equation}
\label{eq:ustar_backbone_state}
\mathbf h_{u,\ell}
=\operatorname{Decoder}_{\theta}
\left(\mathbf C_u,\mathbf s_{<\ell}\right)_{\mathrm{last}}.
\end{equation}
This state summarizes the recent SID history and the generated target prefix. Long-term memories, profile information, and event context enter through the output-side modules described below.  It forms the short-term sequence state used to retrieve
prefix-relevant evidence from the long-term memory banks

\textbf{Prefix-conditioned memory reading.}
The compressed memories are concatenated along the token dimension as
\(\mathbf M_u=[\mathbf Z_{u,1};\mathbf Z_{u,2};\mathbf Z_{u,3}]\).
This step connects the two sequence streams: the recent-prefix state \(\mathbf h_{u,\ell}\) queries the compressed long-term memories through cross-attention. Here, \(\mathbf m_{u,t}\) denotes the \(t\)-th memory token in the long-term memory sequence \(\mathbf M_u\) of user \(u\). For one attention head:
\begin{equation}
\label{eq:ustar_memory_projections}
\begin{aligned}
\mathbf q_{u,\ell}
&=\mathbf W_Q\operatorname{LN}
\left(\mathbf h_{u,\ell}+\mathbf e_{\mathrm{step}}(\ell)\right),\\
\mathbf k_{u,t}
&=\mathbf W_K\mathbf m_{u,t},\\
\mathbf v_{u,t}
&=\mathbf W_V\mathbf m_{u,t}.
\end{aligned}
\end{equation}
The attention weights and retrieved information are then computed as
\begin{equation}
\label{eq:ustar_memory_readout}
\begin{aligned}
a_{u,\ell,t}
&=
\frac{
\exp\left(\mathbf q_{u,\ell}^{\top}\mathbf k_{u,t}/\sqrt{d_a}\right)
}{
\sum_{t'}
\exp\left(\mathbf q_{u,\ell}^{\top}\mathbf k_{u,t'}/\sqrt{d_a}\right)
},\\
\mathbf r_{u,\ell}
&=\sum_t a_{u,\ell,t}\mathbf v_{u,t}.
\end{aligned}
\end{equation}
Attention is normalized over all memory tokens, allowing each decoding step to select information both within and across history views. Because \(\mathbf h_{u,\ell}\) encodes the current prefix \(\mathbf s_{<\ell}\), the attention weights are specific to the current tree node rather than shared across all decoding steps. All three banks remain accessible at every SID position. The multi-head implementation concatenates the head outputs and applies an output projection.

\textbf{Hybrid decoder decision tokens.}
The sequence-evidence path
captures what the user has done at both long-term and recent
horizons. However, the next branch decision may also depend on
who the user is and under which request-time context the deci-
sion is made. This motivates the hybrid design of PATH Decoder,
which combines sequential behavioral evidence with two light-
weight non-sequential decision tokens. Profile-field embeddings are concatenated into \(\mathbf a_u\) and projected into a User Token:
\begin{equation}
\label{eq:ustar_profile_token}
\begin{aligned}
\mathbf U_u
&=\mathbf W_o
\left[
\operatorname{SiLU}(\mathbf W_g\mathbf a_u+\mathbf b_g)
\odot(\mathbf W_v\mathbf a_u+\mathbf b_v)
\right]\\
&\quad+\mathbf b_o.  
\end{aligned}
\end{equation}
Here, \(\mathbf W_g,\mathbf W_v,\mathbf W_o\) and \(\mathbf b_g,\mathbf b_v,\mathbf b_o\) are learnable projection matrices and bias terms, and \(\odot\) denotes element-wise multiplication. Let \(\boldsymbol\tau_u\) denote the event features available before prediction. The corresponding Event Token is
\begin{equation}
\label{eq:ustar_event_representation}
\mathbf e_u^{\mathrm{evt}}
=\mathbf e_{\mathrm{evt}}
+\sum_f\mathbf E_f^{\mathrm{time}}
\left[\tau_u^{(f)}\right].
\end{equation}
Here, \(\mathbf e_{\mathrm{evt}}\) is a learnable base event embedding, \(f\) indexes an event-feature field, \(\tau_u^{(f)}\) is the corresponding discretized feature value, and \(\mathbf E_f^{\mathrm{time}}\) is its embedding table. The User Token provides profile-derived user-bias information, while the Event Token supplies event context. Both participate in the output-side fusion. The final output-side prediction then concatenates the backbone state, retrieved memory, User Token, and Event Token at each decoding step:
\begin{equation}
\label{eq:ustar_prediction_fusion}
\begin{aligned}
\mathbf f_{u,\ell}
&=\operatorname{Concat}
\left(
\mathbf h_{u,\ell},
\mathbf r_{u,\ell},
\mathbf U_u,
\mathbf e_u^{\mathrm{evt}}
\right),\\
\widetilde{\mathbf h}_{u,\ell}
&=\operatorname{MLP}(\mathbf f_{u,\ell}).
\end{aligned}
\end{equation}
The fused representation is projected to the vocabulary at the current SID level:
\begin{equation}
\label{eq:ustar_token_prediction}
\begin{aligned}
\mathbf z_{u,\ell}
&=\mathbf O_\ell\widetilde{\mathbf h}_{u,\ell}+\mathbf b_\ell,\\
\pi_\theta(s_\ell=k\mid x_u,\mathbf s_{<\ell})
&=\operatorname{Softmax}(\mathbf z_{u,\ell})_k.
\end{aligned}
\end{equation}
Here, \(x_u\) denotes the overall user conditions, whereas \(\mathbf C_u\) denotes the SID-only backbone input. The MLP combines prefix-dependent behavioral evidence with profile and event information before token prediction. The compressed memories and their key/value projections can be reused across decoding steps and beams.

\textbf{Supervised fine-tuning (SFT).}
We first train the model with
supervised fine-tuning using next-item maximum likelihood:
\begin{equation}
\label{eq:ustar_sft_objective}
\mathcal L_{\mathrm{SFT}}
=-\mathbb E_{(x_u,v^+)\sim\mathcal D_{\mathrm{SFT}}}
\sum_{\ell=1}^{3}
\log\pi_\theta
\left(s_\ell(v^+)\mid x_u,\mathbf s_{<\ell}(v^+)\right).
\end{equation}
The loss is computed only on the target item's SID; the historical SID tokens serve as context.
\begin{figure}[t]
    \centering
    \includegraphics[width=0.82\columnwidth]{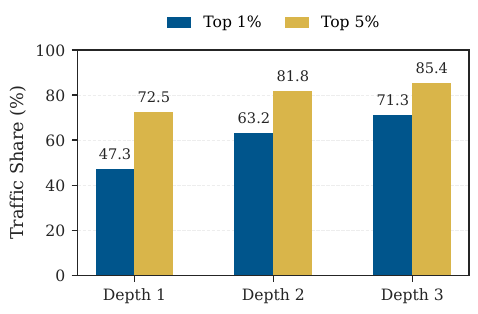}
    \caption{Historical traffic concentration of the top 1\% and top 5\% SID paths across prefix depths.}
    \Description{A grouped bar chart showing top one-percent and top five-percent historical traffic shares at three SID prefix depths.}
    \label{fig:sid_traffic_concentration}
\end{figure}
\subsection{RQ-PIB: Prefix-Balanced Semantic ID
Construction}
\label{sec:rq_pib}
Autoregressive SID decoding makes branch balance a prefix-local issue. At level \(\ell\), a partial SID \(\mathbf s_{<\ell}\) specifies a parent prefix \(p\), and predicting \(s_\ell\) selects one child branch \(k\) under this parent. Thus, the relevant load is the relative historical load among sibling branches under the same prefix, not merely the global usage of a codeword at a level.

Popular SID paths can acquire a frequency-driven advantage in next-token prediction, making it harder for user-specific evidence to influence branch selection. Figure~\ref{fig:sid_traffic_concentration} reveals pronounced concentration in historical path load: the top 1\% of paths account for 47.3\%, 63.17\%, and 71.26\% of the total load at prefix depths one, two, and three, respectively. The corresponding shares for the top 5\% are 72.46\%, 81.79\%, and 85.43\%. Paths are ranked independently at each depth by their aggregated historical load. This concentration motivates mitigating Matthew-effect amplification, in which already popular paths may gain further advantages during model learning.

RQ-PIB, summarized in Algorithm 1, constructs semantic IDs
with prefix-balanced branch organization. It first learns a shared semantic codebook through residual clustering, then rebalances over-
loaded child branches within each parent prefix using minimum-
distortion item moves. This preserves the semantic structure of
residual quantization while reducing local branch dominance dur-
ing autoregressive decoding.

\textbf{Residual semantic clustering}
Let \(\mathbf f_i\) denote the embedding of item \(i\), and let \(\boldsymbol\mu_{\ell,k}\) be codeword \(k\) at level \(\ell\). Given the codes assigned at previous levels, the current residual is

\begin{equation}
\label{eq:residual_quantization}
\mathbf r_i^{(\ell)}
=
\mathbf f_i
-
\sum_{j<\ell}
\boldsymbol\mu_{j,s_j(i)}.
\end{equation}

At the first level, \(\mathbf r_i^{(1)}=\mathbf f_i\). We initialize \(K_\ell\) shared centers with K-Means++ and run standard K-Means on the residuals until convergence, alternating nearest-center assignment and mean-center updates. Let \(z_i\) denote the resulting assignment.  The converged
centers remain fixed during the subsequent prefix-balancing stage,
so later reassignment changes only branch membership, not the
semantic codebook.

\paragraph{Prefix-level load upper bound.}
Let \(w_i\geq0\) denote the historical interaction count of item \(i\),
computed from training-period records. For a parent prefix \(p\) at level
\(\ell\), define
\begin{equation}
\label{eq:ustar_pc_group_safe}
\begin{aligned}
\mathcal I_p&=\{i:\mathbf s_{<\ell}(i)=p\},\\
W_p&=\sum_{i\in\mathcal I_p}w_i,\qquad
m_p=\max_{i\in\mathcal I_p}w_i,\\
W_{p,k}&=\sum_{i\in\mathcal I_p:z_i=k}w_i,
\qquad C_p=\frac{W_p}{K_\ell}.
\end{aligned}
\end{equation}
Here, \(\mathcal I_p\) is the item set under parent prefix \(p\), \(W_p\) is its total load, \(m_p\) is the largest single-item load, \(W_{p,k}\) is the load of child branch \(k\), and \(C_p\) is the average load per child branch.
The average load \(C_p\) gives a natural balancing target, but item weights are indivisible: a single heavy item or a sparse parent group can make a bound based only on \(C_p\) infeasible. We therefore use the effective load upper bound
\begin{equation}
\label{eq:ustar_pc_capacity_safe}
B_p=\max\left\{
\tau_\ell C_p,\;
C_p+\left(1-\frac{1}{K_\ell}\right)m_p
\right\},
\qquad \tau_\ell\geq1,
\end{equation}
and impose \(W_{p,k}\leq B_p\). Here, \(B_p\) is the branch-load upper bound and \(\tau_\ell\) is a level-specific load tolerance. The first term controls the desired tolerance, while the second accounts for item granularity; the bound is computed per parent prefix, with the first level using the empty prefix.

\paragraph{Minimum-distortion branch rebalancing.}
After residual K-Means converges, RQ-PIB freeze the centers and repair
each parent group independently. For a positive-weight item \(i\)
in an overloaded branch \(k\), define
\begin{equation}
\label{eq:ustar_pc_move_safe}
\begin{aligned}
\mathcal Q_i
&=\{q\in\{1,\ldots,K_\ell\}\setminus\{k\}:
W_{p,q}+w_i\leq B_p\},\\
q_i^*
&=\arg\min_{q\in\mathcal Q_i}
\|\mathbf r_i^{(\ell)}-\boldsymbol\mu_{\ell,q}\|_2^2,\\
\Delta_i
&=\|\mathbf r_i^{(\ell)}-\boldsymbol\mu_{\ell,q_i^*}\|_2^2
-\|\mathbf r_i^{(\ell)}-\boldsymbol\mu_{\ell,k}\|_2^2.
\end{aligned}
\end{equation}

Here, \(\mathcal Q_i\) is the feasible destination set, \(q_i^*\) is the closest feasible branch in residual space, and \(\Delta_i\) is the additional reconstruction cost of moving \(i\) from \(k\) to \(q_i^*\).

We relocate the item with the smallest \(\Delta_i\) within the selected
overloaded branch, update the source and destination loads, and
recompute the feasible destinations before the next move. All
\(K_\ell\) codewords, including currently empty branches, are eligible
destinations. The rule minimizes the current feasible move cost, without claiming global optimality or perfectly equal branch loads.

\paragraph{Feasibility and termination.}
Under Eq.~\eqref{eq:ustar_pc_capacity_safe}, every positive-weight item in an
overloaded branch has a feasible destination when all
$K_\ell$ codewords are available and no additional assignment
constraints are imposed. Each move respects the bound and moves
an item at most once, so rebalancing terminates within
$|\mathcal I_p|$ moves; groups with $W_p=0$ or $K_\ell=1$ require
no repair. The greedy procedure claims neither global optimality
nor exact load equality, and does not remove intrinsic item
popularity or guarantee future exposure balance.

\begin{algorithm}[t]
\caption{RQ-PIB: Prefix-Balanced SID Construction}
\label{alg:ustar_pc_rq}
\begin{algorithmic}[1]
\Require Item embeddings $\{\mathbf f_i\}_{i=1}^{N}$,
weights $\{w_i\}$, codebook sizes $\{K_\ell\}_{\ell=1}^{L}$,
tolerances $\{\tau_\ell\}_{\ell=1}^{L}$
\Ensure Prefix-balanced SIDs and shared codebooks

\For{$\ell=1,\ldots,L$}
    \State Compute residuals
    $\mathbf r_i^{(\ell)}
    =\mathbf f_i-\sum_{j<\ell}\boldsymbol\mu_{j,s_j(i)}$
    \State Run K-Means to obtain
    $\{\boldsymbol\mu_{\ell,k}\}$ and assignments $z_i$
    \State Freeze codebook and group items by prefix
    $p=\mathbf s_{<\ell}(i)$

    \ForAll{parent groups $\mathcal I_p$}
        \State Compute branch loads $W_{p,k}$
        and upper bound $B_p$ (Eq.~\ref{eq:ustar_pc_capacity_safe})

        \While{$\exists k: W_{p,k}>B_p$}
            \State Select an overloaded branch $k$
            \State Find feasible moves
            $(i,q)$ satisfying $W_{p,q}+w_i\leq B_p$
            \State Select $(i^*,q^*)$ with minimum
            distortion increase $\Delta_i$
            \State Reassign $z_{i^*}\gets q^*$
            and update branch loads
        \EndWhile
    \EndFor

    \State Set $s_\ell(i)\gets z_i$ for all $i$
\EndFor

\State \Return Item SIDs and shared codebooks
\end{algorithmic}
\end{algorithm}
\subsection{Preference Alignment}
\label{sec:preference_alignment}
Given a recent request prompt $x$, constrained beam search returns
$K$ valid items, which are scored by a frozen ranking model
$r_\phi$. The top-$M$ and bottom-$M$ candidates form all
$M^2$ preference pairs. For an item $v$, let
$\ell_\theta(x,v)$ denote the sum of its SID-token
log-probabilities. Against the frozen SFT reference
$\pi_{\mathrm{ref}}$, we optimize the standard DPO objective:
\begin{equation}
\begin{aligned}
\mathcal L_{\mathrm{DPO}}
=&-\mathbb E_{(x,v^+,v^-)\sim\mathcal D_{\mathrm{pref}}}
\log \sigma \Big(
\beta \big[
\ell_\theta(x,v^+)-\ell_{\mathrm{ref}}(x,v^+)\\
&\qquad\qquad
-\ell_\theta(x,v^-)+\ell_{\mathrm{ref}}(x,v^-)
\big]
\Big).
\end{aligned}
\end{equation}

To preserve supervised next-item generation, we jointly optimize
recent clicked examples:
\begin{equation}
    \mathcal L_{\mathrm{total}}
    =
    \mathcal L_{\mathrm{SFT}}
    +
    \lambda_{\mathrm{DPO}}
    \mathcal L_{\mathrm{DPO}},
\end{equation}
where $\lambda_{\mathrm{DPO}}$ controls the alignment weight.

\subsection{Offline Experiments}
\subsubsection{Experimental Setup}
\label{sec:experimental_setup}

\paragraph{\normalfont\bfseries Dataset and evaluation protocol.}
We use large-scale production logs collected from Youku's homepage
recommendation scenario. SFT uses several months of playback logs,
while daily DPO uses recent request contexts whose beam-search
candidates are scored by a frozen ranking model trained with click
and watch-time feedback. Users are split into training, validation,
and test sets at a ratio of 90:5:5.

\paragraph{\normalfont\bfseries Evaluation metrics.}
We report Hit Ratio at $K$ (HR@$K$) for offline evaluation and
relative improvements over the production Base in Play Conversion
Rate (PCR), Time Spent per User (TS/User), and Total Time Spent
(Total TS) for online evaluation.

\paragraph{\normalfont\bfseries Model and training configuration.}
Unless stated otherwise, we use Qwen3-0.6B as the decoder-only
backbone and represent each item with a fixed three-level RQ-PIB
SID constructed from training-set interaction counts, as described
in Section~\ref{sec:rq_pib}. We train the supervised model with
AdamW using a learning rate of $2\times10^{-5}$. For the preference
alignment stage in Section~\ref{sec:preference_alignment}, we set
$\beta=0.1$ and $\lambda_{\mathrm{DPO}}=0.1$. At inference,
trie-constrained beam search is used to ensure valid SID generation.

\paragraph{\normalfont\bfseries Baselines.}
We compare SPAR with item-space sequential recommenders
GRU4Rec~\cite{hidasi2016gru4rec} and
SASRec~\cite{kang2018sasrec}, and Semantic-ID-based generative
recommenders TIGER~\cite{tiger} and
OneRec~\cite{onerec}.
\begin{figure*}[t]
    \centering
    \includegraphics[width=0.98\textwidth]{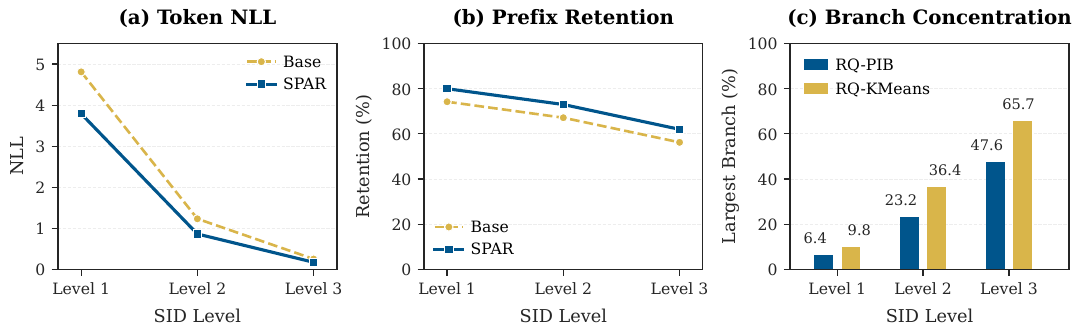}

    \caption{
        SID-level decoding and prefix-conditioned traffic concentration analysis.
        (a) Held-out token NLL.
        (b) Beam-search target-prefix retention.
        (c) Traffic-weighted largest child-branch share.
    }
    \Description{Three plots showing token NLL, beam-search prefix retention, and prefix-conditioned traffic concentration across SID levels.}
    \label{fig:sid_level_analysis}
\end{figure*}
\subsubsection{Evaluation Results}
\begin{table}[t]
    \centering
    \caption{
        Overall retrieval performance.
        The best results are boldfaced, and the second-best
        results are underlined.
    }
    \label{tab:overall_retrieval}

    \begingroup
    \small
    \setlength{\tabcolsep}{4pt}
    \renewcommand{\arraystretch}{1.15}

    \begin{tabularx}{\columnwidth}{
        @{}X
        >{\centering\arraybackslash}X
        >{\centering\arraybackslash}X
        >{\centering\arraybackslash}X
        >{\centering\arraybackslash}X@{}
    }
        \toprule
        Method
        & HR@10
        & HR@20
        & HR@50
        & HR@100 \\
        \midrule

        GRU4Rec
        & 0.3983
        & 0.4951
        & 0.5331
        & 0.5584 \\

        SASRec
        & 0.3630
        & 0.4707
        & 0.5103
        & 0.5381 \\

        TIGER
        & 0.4424
        & 0.5284
        & 0.5674
        & 0.5879 \\

        OneRec
        & \underline{0.4648}
        & \underline{0.5492}
        & \underline{0.5828}
        & \underline{0.6013} \\

        \midrule

        \rowcolor{USTARLightBlue}
        \textbf{SPAR}
        & \textbf{0.4831}
        & \textbf{0.5662}
        & \textbf{0.5947}
        & \textbf{0.6108} \\

        \bottomrule
    \end{tabularx}

    \endgroup
\end{table}
Table~\ref{tab:overall_retrieval} shows that SPAR achieves the best performance across all cutoffs. Compared with the strongest baseline OneRec, HR@10 increases from 0.4648 to 0.4831 and HR@50 from 0.5828 to 0.5947, corresponding to relative gains of 3.94\% and 2.04\%, respectively. Averaged over HR@10/20/50/100, SPAR improves the strongest baseline by 2.66\%. The following ablations isolate the effects of user information, stepwise memory reading, SID construction, and DPO.

\begin{table}[t]
\centering
\caption{Effect of backbone scale on next-item retrieval.}
\label{tab:model_scaling}
\resizebox{\columnwidth}{!}{
\begin{tabular}{lcccc}
\toprule
Backbone
& HR@10
& HR@20
& HR@50
& HR@128 \\
\midrule
Qwen3-0.6B
& 0.4831
& 0.5662
& 0.5947
& 0.6124 \\
Qwen3-1.7B
& 0.4969
& 0.5783
& 0.6060
& 0.6214 \\
Qwen3-4B
& 0.5052
& 0.5856
& 0.6128
& 0.6278 \\
\bottomrule
\end{tabular}
}
\end{table}
\begin{table*}[t]
\centering
\caption{
Ablation study of SPAR.
Part (a) adds individual information sources to Base.
Part (b) compares memory-reading strategies with all
information sources retained.
Part (c) compares SID construction methods while retaining
all information sources and stepwise memory reading.
Part (d) evaluates DPO using the complete SPAR model
as the starting point.
Avg.\ Gain denotes the average absolute improvement over Base
across the four metrics, measured in percentage points. The best result within each ablation block is boldfaced.
}
\label{tab:sft_ablation}

\begingroup
\small
\setlength{\tabcolsep}{6pt}
\renewcommand{\arraystretch}{1.15}

\newcommand{\USTARAbRow}[6]{%
    #1
    & #2
    & #3
    & #4
    & #5
    & #6 \\
}

\newcommand{\USTARAbBoldRow}[6]{%
    \USTARAbRow
        {\textbf{#1}}
        {\textbf{#2}}
        {\textbf{#3}}
        {\textbf{#4}}
        {\textbf{#5}}
        {\textbf{#6}}
}

\begin{tabular*}{\textwidth}{
    @{\extracolsep{\fill}}lccccc@{}
}
    \toprule
    Variant
    & HR@1
    & HR@5
    & HR@10
    & HR@20
    & Avg.\ Gain (pp) \\
    \midrule

    \multicolumn{6}{@{}l}{
        \textit{(a) Contribution of user information}
    } \\
    \addlinespace[3pt]

    \USTARAbRow
        {Base}
        {0.1759}{0.3582}{0.4424}{0.5284}{--}

    \USTARAbRow
        {Base + Event Token}
        {0.1862}{0.3750}{0.4578}{0.5396}{+1.34}

    \USTARAbRow
        {Base + Long-term SID Memories}
        {0.1802}{0.3657}{0.4496}{0.5341}{+0.62}

    \USTARAbRow
        {Base + User Token}
        {0.1806}{0.3661}{0.4502}{0.5347}{+0.67}

    \midrule

    \multicolumn{6}{@{}l}{
        \textit{(b) Memory reading with all information sources}
    } \\
    \addlinespace[3pt]

    \USTARAbRow
        {SPAR (First-step Reuse)}
        {0.1930}{0.3890}{0.4740}{0.5570}{+2.70}

    \USTARAbBoldRow
        {SPAR (Stepwise)}
        {0.1991}{0.3984}{0.4831}{0.5662}{+3.55}

    \midrule

    \multicolumn{6}{@{}l}{
        \textit{(c) SID construction}
    } \\
    \addlinespace[3pt]

    \USTARAbRow
    {SPAR (RQ-KMeans)}
    {0.1794}{0.3549}{0.4320}{0.5385}{0.00}

\USTARAbRow
    {SPAR (Global Capacity-Constrained)}
    {0.1791}{0.3546}{0.4314}{0.5380}{-0.05}

\USTARAbBoldRow
    {SPAR (RQ-PIB)}
    {0.1991}{0.3984}{0.4831}{0.5662}{+3.55}

    \midrule

    \multicolumn{6}{@{}l}{
        \textit{(d) DPO preference alignment}
    } \\
    \addlinespace[3pt]

    \USTARAbRow
        {SPAR (w/o DPO)}
        {0.1991}{0.3984}{0.4831}{0.5662}{+3.55}

    \USTARAbBoldRow
        {SPAR (SFT + DPO)}
        {0.2076}{0.4035}{0.4950}{0.5719}{+4.33}

    \bottomrule
\end{tabular*}

\par\vspace{4pt}
\begin{minipage}{\textwidth}
    \footnotesize

\end{minipage}

\endgroup
\end{table*}
\subsubsection{Analysis of PATH Decoder.}
\label{sec:hierarchical_decoding_analysis}

To examine how SPAR affects individual decoding decisions, we
analyze held-out token negative log-likelihood (NLL) at each SID
level and target-prefix retention during beam search.
These metrics characterize conditional token prediction and
the preservation of target paths during generation, respectively.

\paragraph{\normalfont\bfseries Token prediction across SID levels}
As shown in Figure~\ref{fig:sid_level_analysis}(a), SPAR
consistently reduces held-out token NLL at all three SID levels.
At the first level, NLL decreases from 4.81 to 3.79, corresponding
to a relative reduction of 21.21\%.
At the second and third levels, NLL decreases from 1.23 to 0.86
and from 0.25 to 0.17, yielding relative reductions of 30.08\%
and 32.00\%, respectively.

These improvements indicate that SPAR assigns higher conditional
likelihood to the target tokens throughout the SID hierarchy.
In particular, the reductions at the second and third levels
show improved target-token prediction after the preceding
target prefix has been specified.
The results are consistent with the benefit of incorporating
user information into successive branch decisions.
Our analysis focuses on comparisons between models at the same
SID level, since vocabulary sizes and conditioning contexts
differ across levels.

\paragraph{\normalfont\bfseries Target-path retention during beam search.}
Figure~\ref{fig:sid_level_analysis}(b) examines whether the target
prefix remains in the beam as decoding proceeds.
For Base, the retention rate decreases from 74.23\% at the first
level to 67.15\% at the second level and 56.24\% at the third
level. The latter two transitions correspond to decreases of
7.08 and 10.91 percentage points, respectively, illustrating
the progressive loss of target paths during generation.

Prefix retention complements the teacher-forced NLL analysis:
NLL evaluates the next-token prediction given a correct prefix,
whereas retention measures whether the target path remains
available under the model's actual decoding decisions.
Together, these measurements provide a basis for examining
whether improvements in conditional prediction translate into
better preservation of target paths.

%

\subsubsection{Ablation Study}
\label{sec:sft_ablation_analysis}

Table 3 groups ablations by user information,
memory-reading strategy, SID construction, and preference
alignment. All variants in Parts (a)--(c) are trained with SFT
only and do not use DPO. Part (d) starts from the complete SPAR
SFT checkpoint and evaluates the additional effect of DPO while
keeping the SID mapping, model architecture, and decoding
configuration unchanged.

\paragraph{\normalfont\bfseries Contribution of user information.}
Part (a) shows that each information source improves all reported cutoffs. Event Token gives the largest individual average gain (+1.34 pp), while long-term SID memories and User Token add +0.62 and +0.67 pp. Combining all sources in SPAR yields +3.55 pp on average over Base.

\paragraph{\normalfont\bfseries Effect of PATH Decoder.}
Part (b) shows that replacing PATH Decoder with First-step Reuse reduces the average gain from +3.55 to +2.70 pp, isolating the value of prefix-adaptive memory access.

\paragraph{\normalfont\bfseries Effect of RQ-PIB}
Part (c) evaluates the proposed Prefix-Capacitated SID Construction by comparing three strategies: ordinary RQ-KMeans, which performs residual semantic clustering without load constraints; Global Capacity-Constrained RQ-KMeans, which balances codeword loads independently of preceding prefixes; and RQ-PIB, which balances child-branch loads within each parent prefix to align with autoregressive decoding. All variants share identical user information, model configurations, and training settings.

As shown in Table 3, global capacity constraints yield slightly lower performance than ordinary RQ-KMeans, suggesting that balancing aggregate codeword loads alone is insufficient. In contrast, RQ-PIB consistently outperforms both baselines, achieving absolute gains of 2.00, 4.38, 5.17, and 2.82 percentage points over the global variant on HR@1, HR@5, HR@10, and HR@20, respectively. These results highlight the importance of prefix-conditioned load balancing in SID construction.

\paragraph{\normalfont\bfseries Effect of DPO preference alignment.}
Part (d) of Table~\ref{tab:sft_ablation} compares the complete
SPAR model with its DPO-aligned counterpart.
Both variants use the same SID mapping, model architecture,
and decoding configuration, with DPO applied to the complete
SFT checkpoint.

DPO improves HR@1, HR@5, HR@10, and HR@20 from
0.1991, 0.3984, 0.4831, and 0.5662 to
0.2076, 0.4035, 0.4950, and 0.5719, respectively.
These improvements correspond to relative gains of
4.27\%, 1.28\%, 2.46\%, and 1.01\%, with an average
absolute improvement of 0.78 percentage points over
the SFT model.
The largest relative improvement is observed at HR@1,
while positive gains are maintained at all larger cutoffs.
These results support the additional retrieval benefits
of DPO alignment on top of prefix-capacitated SID construction
and stepwise personalized decoding.
\subsubsection{Effect on prefix-conditioned traffic concentration.}
Figure~\ref{fig:sid_level_analysis}(c) compares the traffic-weighted
largest child-branch share at each SID level under ordinary
RQ-KMeans and RQ-PIB. RQ-PIB reduces this share from 9.8\% to 6.4\% at Level 1,
from 36.4\% to 23.2\% at Level 2, and from 65.7\% to 47.6\%
at Level 3.
These changes correspond to absolute reductions of 3.4, 13.2,
and 18.1 percentage points, or relative reductions of 34.69\%,
36.26\%, and 27.55\%, respectively.
The consistently lower values indicate that, on a
traffic-weighted basis, the constructed hierarchy is less
dominated by its highest-load child branches.
The reductions at Levels 2 and 3 further show that this
improvement extends beyond the root to branching decisions
conditioned on single-token and two-token prefixes.
These results support the effectiveness of our SID construction
strategy in controlling prefix-conditioned load concentration
at the branching points used by autoregressive decoding.

\subsubsection{Scaling with Model Size}
\label{sec:model_scaling}

Table 2 studies Qwen3 backbones
with 0.6B, 1.7B, and 4B parameters under the same data, SID code-
books, long-history encoder, and evaluation protocol. Larger back-
bones consistently improve retrieval, but the gain from 1.7B to 4B is
smaller than from 0.6B to 1.7B, suggesting diminishing returns and
making 1.7B a favorable quality-capacity trade-off in our setting.


\subsection{Online A/B Tests}
\begin{table}[t]
    \centering
    \caption{
        Online A/B test results in Youku's homepage recommendation
        scenario. Values indicate relative improvements over the
        production Base model.
    }
    \label{tab:online_ab}
    \begin{tabular}{lc}
        \toprule
        Metric & Relative Improvement \\
        \midrule
        Play Conversion Rate & +3.38\% \\
        Time Spent per User   & +3.95\% \\
        Total TS Duration     & +3.98\% \\
        \bottomrule
    \end{tabular}
\end{table}
\label{sec:online_ab}

We deploy SPAR in Youku's production recommendation system and
conduct an online A/B test in the homepage scenario. The experiment
consists of two user buckets, each receiving approximately 10\% of the
total traffic. The control bucket is served by the production Base
model, while the treatment bucket deploys the complete SPAR model
after SFT and DPO alignment. As shown in Table~\ref{tab:online_ab},
SPAR improves Play Conversion Rate by 3.38\%, TS Value per User by
3.95\%, and Total TS Duration by 3.98\% over the Base model. The
simultaneous improvements in playback conversion and time-spent
metrics indicate that SPAR not only increases the likelihood that a
recommended item triggers playback, but also improves the subsequent
watch-time outcome, demonstrating its effectiveness in Youku's
production environment.

\section{Conclusion}
\label{sec:conclusion}
In this paper, we presented SPAR, an industrial generative recommendation framework combining RQ-PIB and PATH Decoder to address popularity-driven branch imbalance and insufficient stepwise personalization. RQ-PIB balances historical interaction loads among sibling branches while preserving semantic reconstruction quality, mitigating local branch concentration. PATH Decoder retrieves relevant long-term user memories conditioned on the current decoding state and generated prefix, integrating them with recent behaviors, user profiles, and event context to guide each SID token decision. Together, these components improve branch organization and stepwise personalized decoding.

Offline experiments on Youku's production recommendation data show that SPAR outperforms the strongest baseline, achieving an average relative improvement of 2.66\% across HR@10/20/50/100. Further analysis confirms improvements in target-prefix retention and historical traffic balance. A 7-day online A/B test on 10\% of Youku's production traffic further demonstrates gains of 3.95\% in average watch time per user and 3.38\% in play conversion rate. These results demonstrate the effectiveness of SPAR in improving retrieval accuracy, hierarchical decoding reliability, and user engagement in large-scale generative recommendation.
\section{Acknowledgments}

This work was supported by HUJING Digital Media \&
Entertainment Group through the HUJING Digital Media \&
Entertainment Research Intern Program, the STI 2030---Major
Projects (Grant No. 2024ZD0523903), and the National Natural
Science Foundation of China (Grant Nos. T2293771 and
42361144718).

\bibliographystyle{ACM-Reference-Format}
\bibliography{sample-base}

@article{tiger,
  title={Recommender systems with generative retrieval},
  author={Rajput, Shashank and Mehta, Nikhil and Singh, Anima and Hulikal Keshavan, Raghunandan and Vu, Trung and Heldt, Lukasz and Hong, Lichan and Tay, Yi and Tran, Vinh and Samost, Jonah and others},
  journal={Advances in Neural Information Processing Systems},
  volume={36},
  pages={10299--10315},
  year={2023}
}

@article{onerec,
  title={Onerec: Unifying retrieve and rank with generative recommender and iterative preference alignment},
  author={Deng, Jiaxin and Wang, Shiyao and Cai, Kuo and Ren, Lejian and Hu, Qigen and Ding, Weifeng and Luo, Qiang and Zhou, Guorui},
  journal={arXiv preprint arXiv:2502.18965},
  year={2025}
}

@misc{onerecv2,
  title        = {OneRec-V2 Technical Report},
  author       = {Guorui Zhou and Hengrui Hu and Hongtao Cheng and Huanjie Wang
                  and Jiaxin Deng and Jinghao Zhang and Kuo Cai and Lejian Ren
                  and Lu Ren and Liao Yu and Pengfei Zheng and Qiang Luo
                  and Qianqian Wang and Qigen Hu and Rui Huang and Ruiming Tang
                  and Shiyao Wang and Shujie Yang and Tao Wu and Wuchao Li
                  and Xinchen Luo and Xingmei Wang and Yi Su and Yunfan Wu
                  and Zexuan Cheng and Zhanyu Liu and Zixing Zhang and Bin Zhang
                  and Boxuan Wang and Chaoyi Ma and Chengru Song and Chenhui Wang
                  and Chenglong Chu and Di Wang and Dongxue Meng and Dunju Zang
                  and Fan Yang and Fangyu Zhang and Feng Jiang and Fuxing Zhang
                  and Gang Wang and Guowang Zhang and Han Li and Honghui Bao
                  and Hongyang Cao and Jiaming Huang and Jiapeng Chen
                  and Jiaqiang Liu and Jinghui Jia and Kun Gai and Lantao Hu
                  and Liang Zeng and Qiang Wang and Qidong Zhou and Rongzhou Zhang
                  and Shengzhe Wang and Shihui He and Shuang Yang and Siyang Mao
                  and Sui Huang and Tiantian He and Tingting Gao and Wei Yuan
                  and Xiao Liang and Xiaoxiao Xu and Xugang Liu and Yan Wang
                  and Yang Zhou and Yi Wang and Yiwu Liu and Yue Song
                  and Yufei Zhang and Yunfeng Zhao and Zhixin Ling and Ziming Li},
  year         = {2025},
  eprint       = {2508.20900},
  archivePrefix = {arXiv},
  primaryClass = {cs.IR}
}

@article{unirec,
  title={UniRec: Bridging the Expressive Gap between Generative and Discriminative Recommendation via Chain-of-Attribute},
  author={Wang, Ziliang and Lin, Gaoyun and Wang, Xuesi and Liang, Shaoqiang and Huang, Yili and Bian, Weijie and Zhang, Li and Cai, Mingchen and Dong, Jian and Zhang, Guanxing},
  journal={arXiv preprint arXiv:2604.12234},
  year={2026}
}

@inproceedings{youtubednn,
  title={Deep neural networks for youtube recommendations},
  author={Covington, Paul and Adams, Jay and Sargin, Emre},
  booktitle={Proceedings of the 10th ACM conference on recommender systems},
  pages={191--198},
  year={2016}
}

@inproceedings{kdd,
  title={Stop treating collisions equally: Qualification-aware semantic id learning for recommendation at industrial scale},
  author={Hu, Zheng and Chen, Yuxin and Pan, Yongsen and Yuan, Xu and Yin, Yuting and Wang, Daoyuan and Xia, Boyang and Luo, Zefei and Wang, Hongyang and Ni, Songhao and others},
  booktitle={Proceedings of the 32nd ACM SIGKDD Conference on Knowledge Discovery and Data Mining V. 2},
  pages={7401--7411},
  year={2026}
}

@inproceedings{geng2022p5,
  author    = {Geng, Shijie and Liu, Shuchang and Fu, Zuohui and Ge, Yingqiang and Zhang, Yongfeng},
  title     = {Recommendation as Language Processing ({RLP}): A Unified Pretrain, Personalized Prompt \& Predict Paradigm ({P5})},
  booktitle = {Proceedings of the 16th ACM Conference on Recommender Systems},
  year      = {2022},
  pages     = {299--315},
  publisher = {ACM},
  doi       = {10.1145/3523227.3546767}
}

@inproceedings{zhai2024hstu,
  author    = {Zhai, Jiaqi and Liao, Lucy and Liu, Xing and Wang, Yueming and Li, Rui and Cao, Xuan and Gao, Leon and Gong, Zhaojie and Gu, Fangda and He, Jiayuan and Lu, Yinghai and Shi, Yu},
  title     = {Actions Speak Louder than Words: Trillion-Parameter Sequential Transducers for Generative Recommendations},
  booktitle = {Proceedings of the 41st International Conference on Machine Learning},
  series    = {Proceedings of Machine Learning Research},
  volume    = {235},
  year      = {2024},
  pages     = {58484--58509},
  publisher = {PMLR},
  url       = {https://proceedings.mlr.press/v235/zhai24a.html}
}

@inproceedings{hidasi2016gru4rec,
  author    = {Hidasi, Bal{\'a}zs and Karatzoglou, Alexandros and Baltrunas, Linas and Tikk, Domonkos},
  title     = {Session-based Recommendations with Recurrent Neural Networks},
  booktitle = {International Conference on Learning Representations},
  year      = {2016},
  url       = {https://arxiv.org/abs/1511.06939}
}

@inproceedings{kang2018sasrec,
  author    = {Kang, Wang-Cheng and McAuley, Julian},
  title     = {Self-Attentive Sequential Recommendation},
  booktitle = {2018 IEEE International Conference on Data Mining (ICDM)},
  year      = {2018},
  pages     = {197--206},
  publisher = {IEEE},
  doi       = {10.1109/ICDM.2018.00035}
}

@inproceedings{li2019mind,
  author    = {Li, Chao and Liu, Zhiyuan and Wu, Mengmeng and Xu, Yuchi and Zhao, Huan and Huang, Pipei and Kang, Guoliang and Chen, Qiwei and Li, Wei and Lee, Dik Lun},
  title     = {Multi-Interest Network with Dynamic Routing for Recommendation at Tmall},
  booktitle = {Proceedings of the 28th ACM International Conference on Information and Knowledge Management},
  year      = {2019},
  pages     = {2615--2623},
  publisher = {ACM},
  doi       = {10.1145/3357384.3357814}
}

@inproceedings{zhou2018din,
  author    = {Zhou, Guorui and Song, Chengru and Zhu, Xiaoqiang and Fan, Ying and Zhu, Han and Ma, Xiao and Yan, Yanghui and Jin, Junqi and Li, Han and Gai, Kun},
  title     = {Deep Interest Network for Click-Through Rate Prediction},
  booktitle = {Proceedings of the 24th ACM SIGKDD International Conference on Knowledge Discovery \& Data Mining},
  year      = {2018},
  pages     = {1059--1068},
  publisher = {ACM},
  doi       = {10.1145/3219819.3219823}
}

@inproceedings{zhou2019dien,
  author    = {Zhou, Guorui and Mou, Na and Fan, Ying and Pi, Qi and Bian, Weijie and Zhou, Chang and Zhu, Xiaoqiang and Gai, Kun},
  title     = {Deep Interest Evolution Network for Click-Through Rate Prediction},
  booktitle = {Proceedings of the AAAI Conference on Artificial Intelligence},
  volume    = {33},
  number    = {01},
  year      = {2019},
  pages     = {5941--5948},
  doi       = {10.1609/aaai.v33i01.33015941}
}

@inproceedings{pi2019mimn,
  author    = {Pi, Qi and Bian, Weijie and Zhou, Guorui and Zhu, Xiaoqiang and Gai, Kun},
  title     = {Practice on Long Sequential User Behavior Modeling for Click-Through Rate Prediction},
  booktitle = {Proceedings of the 25th ACM SIGKDD International Conference on Knowledge Discovery \& Data Mining},
  year      = {2019},
  pages     = {2671--2679},
  publisher = {ACM},
  doi       = {10.1145/3292500.3330666}
}

@inproceedings{pi2020sim,
  author    = {Pi, Qi and Zhou, Guorui and Zhang, Yujing and Wang, Zhe and Ren, Lejian and Fan, Ying and Zhu, Xiaoqiang and Gai, Kun},
  title     = {Search-based User Interest Modeling with Lifelong Sequential Behavior Data for Click-Through Rate Prediction},
  booktitle = {Proceedings of the 29th ACM International Conference on Information \& Knowledge Management},
  year      = {2020},
  pages     = {2685--2692},
  publisher = {ACM},
  doi       = {10.1145/3340531.3412744}
}

@inproceedings{wang2024letter,
  author    = {Wang, Wenjie and Bao, Honghui and Lin, Xinyu and Zhang, Jizhi and Li, Yongqi and Feng, Fuli and Ng, See-Kiong and Chua, Tat-Seng},
  title     = {Learnable Item Tokenization for Generative Recommendation},
  booktitle = {Proceedings of the 33rd ACM International Conference on Information and Knowledge Management},
  year      = {2024},
  pages     = {2400--2409},
  publisher = {ACM},
  doi       = {10.1145/3627673.3679569}
}

@inproceedings{zhu2024cost,
  title={Cost: Contrastive quantization based semantic tokenization for generative recommendation},
  author={Zhu, Jieming and Jin, Mengqun and Liu, Qijiong and Qiu, Zexuan and Dong, Zhenhua and Li, Xiu},
  booktitle={Proceedings of the 18th ACM Conference on Recommender Systems},
  pages={969--974},
  year={2024}
}

@inproceedings{hou2025generating,
  title={Generating long semantic ids in parallel for recommendation},
  author={Hou, Yupeng and Li, Jiacheng and Shin, Ashley and Jeon, Jinsung and Santhanam, Abhishek and Shao, Wei and Hassani, Kaveh and Yao, Ning and McAuley, Julian},
  booktitle={Proceedings of the 31st ACM SIGKDD Conference on Knowledge Discovery and Data Mining V. 2},
  pages={956--966},
  year={2025}
}

@inproceedings{sun2019bert4rec,
  title={BERT4Rec: Sequential recommendation with bidirectional encoder representations from transformer},
  author={Sun, Fei and Liu, Jun and Wu, Jian and Pei, Changhua and Lin, Xiao and Ou, Wenwu and Jiang, Peng},
  booktitle={Proceedings of the 28th ACM international conference on information and knowledge management},
  pages={1441--1450},
  year={2019}
}

@inproceedings{bao2023tallrec,
  title={Tallrec: An effective and efficient tuning framework to align large language model with recommendation},
  author={Bao, Keqin and Zhang, Jizhi and Zhang, Yang and Wang, Wenjie and Feng, Fuli and He, Xiangnan},
  booktitle={Proceedings of the 17th ACM conference on recommender systems},
  pages={1007--1014},
  year={2023}
}

@inproceedings{li2023text,
  title={Text is all you need: Learning language representations for sequential recommendation},
  author={Li, Jiacheng and Wang, Ming and Li, Jin and Fu, Jinmiao and Shen, Xin and Shang, Jingbo and McAuley, Julian},
  booktitle={Proceedings of the 29th ACM SIGKDD conference on knowledge discovery and data mining},
  pages={1258--1267},
  year={2023}
}

@inproceedings{zheng2024adapting,
  title={Adapting large language models by integrating collaborative semantics for recommendation},
  author={Zheng, Bowen and Hou, Yupeng and Lu, Hongyu and Chen, Yu and Zhao, Wayne Xin and Chen, Ming and Wen, Ji-Rong},
  booktitle={2024 IEEE 40th International Conference on Data Engineering (ICDE)},
  pages={1435--1448},
  year={2024},
  organization={IEEE}
}

@inproceedings{li2024calrec,
  title={Calrec: Contrastive alignment of generative llms for sequential recommendation},
  author={Li, Yaoyiran and Zhai, Xiang and Alzantot, Moustafa and Yu, Keyi and Vuli{\'c}, Ivan and Korhonen, Anna and Hammad, Mohamed},
  booktitle={Proceedings of the 18th ACM Conference on Recommender Systems},
  pages={422--432},
  year={2024}
}

@inproceedings{hou2022towards,
  title={Towards universal sequence representation learning for recommender systems},
  author={Hou, Yupeng and Mu, Shanlei and Zhao, Wayne Xin and Li, Yaliang and Ding, Bolin and Wen, Ji-Rong},
  booktitle={Proceedings of the 28th ACM SIGKDD conference on knowledge discovery and data mining},
  pages={585--593},
  year={2022}
}

@inproceedings{zhou2020s3rec,
  title={S3-rec: Self-supervised learning for sequential recommendation with mutual information maximization},
  author={Zhou, Kun and Wang, Hui and Zhao, Wayne Xin and Zhu, Yutao and Wang, Sirui and Zhang, Fuzheng and Wang, Zhongyuan and Wen, Ji-Rong},
  booktitle={Proceedings of the 29th ACM international conference on information \& knowledge management},
  pages={1893--1902},
  year={2020}
}

@inproceedings{li2023blip,
  title={Blip-2: Bootstrapping language-image pre-training with frozen image encoders and large language models},
  author={Li, Junnan and Li, Dongxu and Savarese, Silvio and Hoi, Steven},
  booktitle={International conference on machine learning},
  pages={19730--19742},
  year={2023},
  organization={PmLR}
}

@inproceedings{ripor,
  title={Scalable and effective generative information retrieval},
  author={Zeng, Hansi and Luo, Chen and Jin, Bowen and Sarwar, Sheikh Muhammad and Wei, Tianxin and Zamani, Hamed},
  booktitle={Proceedings of the ACM Web Conference 2024},
  pages={1441--1452},
  year={2024}
}

@inproceedings{ed2,
  title={Unleash llms potential for sequential recommendation by coordinating dual dynamic index mechanism},
  author={Yin, Jun and Zeng, Zhengxin and Li, Mingzheng and Yan, Hao and Li, Chaozhuo and Han, Weihao and Zhang, Jianjin and Liu, Ruochen and Sun, Hao and Deng, Weiwei and others},
  booktitle={Proceedings of the ACM on Web Conference 2025},
  pages={216--227},
  year={2025}
}

@inproceedings{categorytree,
  title={Learning category trees for ID-based recommendation: exploring the power of differentiable vector quantization},
  author={Liu, Qijiong and Xiao, Jiaren and Fan, Lu and Zhu, Jieming and Wu, Xiao-Ming},
  booktitle={Proceedings of the ACM Web Conference 2024},
  pages={3521--3532},
  year={2024}
}

@inproceedings{cllm4rec,
  title={Collaborative large language model for recommender systems},
  author={Zhu, Yaochen and Wu, Liang and Guo, Qi and Hong, Liangjie and Li, Jundong},
  booktitle={Proceedings of the ACM Web Conference 2024},
  pages={3162--3172},
  year={2024}
}

@inproceedings{rella,
  title={Rella: Retrieval-enhanced large language models for lifelong sequential behavior comprehension in recommendation},
  author={Lin, Jianghao and Shan, Rong and Zhu, Chenxu and Du, Kounianhua and Chen, Bo and Quan, Shigang and Tang, Ruiming and Yu, Yong and Zhang, Weinan},
  booktitle={Proceedings of the ACM Web Conference 2024},
  pages={3497--3508},
  year={2024}
}

@inproceedings{retr,
  title={Recommender transformers with behavior pathways},
  author={Yao, Zhiyu and Chen, Xinyang and Wang, Sinan and Dai, Qinyan and Li, Yumeng and Zhu, Tanchao and Long, Mingsheng},
  booktitle={Proceedings of the ACM Web Conference 2024},
  pages={3643--3654},
  year={2024}
}

\end{document}